\documentclass[twocolumn,showpacs,preprintnumbers,amsmath,amssymb,superscriptaddress]{revtex4}

\usepackage{graphicx}
\usepackage{dcolumn}
\usepackage{bm}
\usepackage{longtable}

\usepackage{color}
\usepackage{latexsym}
\usepackage{amssymb}
\usepackage{amsfonts}
\usepackage{amsmath}

\usepackage[latin1]{inputenc}
\usepackage[english]{babel}

\def\be{\begin{equation}}
\def\ee{\end{equation}}
\def\bc{\begin{center}}
\def\ec{\end{center}}

\newtheorem{theorem}{Theorem}

\newtheorem{remark}[theorem]{Remark}
\newtheorem{definition}[theorem]{Definition}
\newtheorem{proposition}[theorem]{Proposition}

\begin{document}


\title{Effective interactions in group competition with strategic diffusive dynamics}

\author{Elena Agliari}
\affiliation{Dipartimento di Fisica and INFN, Universit\`a di Parma, Italy}
\author{Adriano Barra}
\affiliation{Dipartimento di Fisica, Sapienza Universit\`a di Roma, Italy}
\affiliation{Dipartimento di Matematica, Universit\`a di Bologna, Italy}
\author{Raffaella Burioni}
\affiliation{Dipartimento di Fisica and INFN, Universit\`a di Parma, Italy}
\author{Federico Camboni}
\affiliation{Dipartimento di Fisica, Sapienza Universit\`a di Roma, Italy}
\author{Pierluigi Contucci} 
\affiliation{Dipartimento di Matematica, Universit\`a di Bologna, Italy}

\date{\today}

\begin{abstract}
We analyze, on a random graph, a {\em diffusive
strategic dynamics} with pairwise interactions, where nor Glauber
prescription, neither detailed balance hold. We observe numerically
that such a dynamics reaches a well
defined steady state that fulfills a {\it shift} property: the critical
temperature of the canonical ferromagnetic phase transition is higher with respect to the
expected equilibrium one, known both numerically via Glauber
relaxation or Monte Carlo simulations as well as analytically via
cavity techniques or replica approaches.
\newline
We show how the relaxed states of this kind of dynamics can be
described by statistical mechanics equilibria of a diluted p-spin
model, for a suitable non-integer real $p$. Several implications
from both theoretical physics and quantitative sociology points of view
are discussed.
\end{abstract}

%
\maketitle
\section{Introduction}

Born as a theoretical background for thermodynamics,
statistical mechanics provides nowadays a flexible approach
to several scientific problems whose depth and wideness
increases continuously. In the last decades in fact
complex systems statistical mechanics has invaded
fields as diverse as spin glasses \cite{MPV}, neural
networks \cite{amit}, protein folding \cite{huang},
immunological memory \cite{immune}, and also made some
attempt to describe social networks \cite{PL1},
theoretical economy \cite{ton} and urban planning \cite{intro2}.
\newline
In this paper we study statistical mechanics
of random diluted mean field systems, paying particular attention
to its applications in social science: in a nutshell we show that
the relaxed states of a particular non-Glauber two-body dynamics
\cite{prima,ABCV2005} can be ''effectively'' described by a
diluted p-spin model for a suitable non-integer real $p$.
\newline
After a review of the {\em diffusive strategic dynamics}
previously introduced in \cite{BBCV2002,agliari2,agliari3}, we implement it
on an Erd\"{o}s-Renyi mean field random model and work out
the corresponding mean field equilibria: each agent (spin) is selected through a diffusive rule,
and its flipping probability  is not weighed ``a la Glauber" \cite{amit}.
The probabilities of its nearest neighbor flips are
selected by favouring the flips which produce the maximum
energy gain. We stress that this operation, although dynamically pairwise,
effectively involves more than pairwise information evaluation, as the chosen agent
interacts both with the first selected one as well as
with its nearest neighbors, as a whole.
\newline
This dynamics is shown to relax to a well defined steady state,
where all the properties of stationarity are recovered
\cite{liquid}. However a peculiarity happens, with respect to the
standard relaxation investigated both by means of Glauber
dynamics \cite{amit}, and well known both by Montecarlo simulation
and analytical results \cite{ABC}: the critical parameters
(temperature, or equivalently the inverse of the strength of the interactions)
are higher, of a few percent, than the expected.
\newline
The whole scenario suggests a ``hidden'' many body interaction, encoded into the
particular rule for selecting the spins. To investigate this
feature we perform further numerical analysis, which strongly
supports a more-than-two body effective interaction. Then we
work out analytically a theory for the randomly diluted p-spin
model so to fit an effective $p\in \mathbb{R}$, which turns out to
be $p=2.15$, in order to match the numerical data available by the
dynamics. This result has implication both in theoretical physics,
as well as in quantitative sociology, where the effective interactions
always play an important role in decision making \cite{durlauf,mcfadden}.
\newline
The paper is organized as follows:
\newline
In section \ref{uno} our dynamics is introduced and shortly
discussed; then further numerical investigations toward a better
understanding of a $p>2$ behavior are presented.
\newline
In section \ref{tre} the randomly diluted p-spin model is defined
and exploited in all details, both analytically within the cavity
field techniques as well as numerically, within a Montecarlo
approach. Full agreement is found among the two methods.
\newline
At the end, the last section is left for conclusions: the
effective interaction is found and its implications analyzed.
\newline
Furthermore, even though the paper is written within a theoretical
physics approach, remarks concerning the application to
quantitative sociology are sparse through all the work.
\newline
As a last remark, for a better reading of the manuscript, we decided to put all the long proofs
of the various theorems in the appendix.

\section{Diffusive strategic dynamics}\label{uno}
In this section we shortly discuss the general approach for the
numerical investigation of an Ising-like system \cite{ABCV2005},
then we introduce a particular dynamics and we explain the
motivation behind our choice.

\subsection{A strategy avoiding Glauber prescription}

In order to simulate the dynamical evolution of a system described
by a Curie-Weiss Hamiltonian $\hat{H}_N(\sigma)$, living on an
Erd\"{o}s-Renyi random graph \cite{ABC,gt2},
$$
\hat{H}_N(\sigma)= \sum_{ij}^{N}A_{ij}\sigma_{i}\sigma_{j},
\label{IsingER}
$$
where $A_{ij}$ is a Poisson random adjacency matrix that, on
average, connects $\alpha_i$ sites to the generic $i^{th}$ spin
(and thermodynamically $\alpha_i \to \alpha$, the connectivity,
whenever $N \to \infty$), several different algorithms have been
introduced. Among them a well established one is the so-called
single-flip algorithm, which makes the system evolve by means of
successive spin-flips, where we call ``flip'' on the node $j$ the
transformation $s_j \rightarrow -s_j$ \cite{liggett}.

More precisely, the generic single-flip algorithm is made up of
two parts: first we need a rule according to which we select a
spin to be updated, then we need a probability distribution which
states how likely the spin-flip is. As for the latter, following
the Glauber rule, given a configuration $\mathbf{s}$, the
probability for the spin-flip on the $j$-th node reads off as
\begin{equation} \label{eq:Glauber}
p(\textbf{s},j,\mathbf{J}) = \frac{1}{1 + e^{\displaystyle
\beta\Delta H(\textbf{s},j,\mathbf{J})  }},
\end{equation}
where $\Delta H(\textbf{s},j,\mathbf{J})= 2\sigma_i\sum_j
A_{ij}\sigma_j$ is the variation in the cost function due to the
flip $s_j \rightarrow -s_j.$ Hence, for single-flip dynamics the
cost variation $\Delta H$, consequent to a flip, only depends on
the spin of a few sites, viz. the $j$-th one undergoing the
flipping process and its $\alpha_j$ nearest-neighbors (for the
sake of clearness, we remember that the Erd\"{o}s-Renyi, even if the
amount of nearest neighbors is finite, still allows the model
to be analyzed via mean-field techniques).
\newline
As for the selection rule according to which sites are extracted,
there exist several different choices, ranging from purely random to deterministic. In several contexts
(condensed-matter physics \cite{BBCV2002}, sociology \cite{prima}
etc.) unless no peculiar mechanisms or strategies are at work, the
random updating  seems to be the most plausible.
In this case the probability that the current configuration
$\mathbf{s}$ changes into $\mathbf{s}_j'$ due to the flip $s_j
\rightarrow - s_j$, reads off as
\begin{equation}\label{eq:prob_random}
\mathcal{P}^{\mathcal{R}} (\mathbf{s},j;\mathbf{J}) = \frac{1}{N}
p(\mathbf{s},j,\mathbf{J}).
\end{equation}
This algorithm mimics the coupling between the magnetic system
with the thermal vibrations of the underlying structure, usually
meant as heat-bath. The dynamics generated by
$\mathcal{P}^{\mathcal{R}}$ has been intensively studied in the
past (see e.g. \cite{barkema}) and it has been shown to lead the
system to the canonical equilibrium distribution, derived
form the cost function $H_N(\sigma)$.

However, there exist several other different mechanisms yielding
single flips \cite{kcm}. For example, we can think of a system
endowed with a local thermostat where diffusing excitations affect the spin dynamics.
Then it is reasonable to suppose that the spin-flips, and the
related energy changes, have a diffusive character and that such a
diffusion is biased towards those regions of the sample where
energy variations are more likely to occur. Also in a social
context a spin-flip can occur as a result of a direct interaction
(phone call, mail exchange, etc.) between two neighbors and if
agent $i$ has just undergone an opinion-flip he will, in turn,
prompt one out of his $\alpha_i$ neighbors to change opinion
(opinion in social context plays the role of the spin orientation
in material systems).

These aspects are neglected by traditional dynamics and can not be
described by a random updating rule.
In the past a different relaxation dynamics  has been introduced and it is able to take into account these aspects, namely:\\
\textit{i}. the selection rule exhibits a \textit{diffusive
character}: The sequence of sites selected for the updating can be
thought of as the  path of a random walk moving on the
underlying structure.\\ \textit{ii}. the diffusion is
\textit{biased}: The $\alpha_i$ neighbors are not equally likely
to be chosen but, amongst the $\alpha_i$ neighbors, the most
likely to be selected is also the most likely to undergo a
spin-flip, namely the one which minimizes $\Delta
H(\mathbf{s},j,\mathbf{J})$.

Let us now formalize how the dynamics  works. Our MC simulations
are made up of successive steps \cite{prima}:

- Being $i$ the newest updated spin/agent (at the very first step
$i$ is extracted randomly from the whole set of agents), we
consider the corresponding set of nearest-neighbors defined as
$\mathcal{N}_i=\{ i_1, i_2, ..., i_{\alpha_i}\}$; we possibly
consider also the subset $\tilde{\mathcal{N}}_i \subseteq
\mathcal{N}_i$ whose elements are nearest- neighbors of $i$ not
sharing the same orientation/opinion: $j \in \tilde{\mathcal{N}}_i
\Leftrightarrow j \in \mathcal{N}_i \wedge s_i s_j = -1$. Now, for
any $j \in \mathcal{N}_i$ we compute the cost function variation
$\Delta H (\mathbf{s},j,\mathbf{J})$, which would result if the
flip $s_j \rightarrow -s_j$ occurred; notice that $\Delta H
(\mathbf{s},j,\mathbf{J})$ involves not only the nearest-neighbors
of $i$.

- We calculate the probability of opinion-flip for all the nodes
in $\mathcal{N}_i$, hence obtaining
$p(\textbf{s},i_1,\mathbf{J}),p(\textbf{s},i_2,\mathbf{J}),...,p(\textbf{s},i_{\alpha_i},\mathbf{J})$,
where $p(\textbf{s},s_j',\mathbf{J})$ (see Eq.~\ref{eq:Glauber}),
is the probability that the current configuration $\textbf{s}$
changes due to a flip on the $j$-th site.

- We calculate the probability ${\cal P}^{\mathcal{S}}
(\textbf{s};i,j;\mathbf{J})$ that among all possible $\alpha_i$
opinion-flips considered just the $j$-th one is realized; this is
obtained by properly normalizing the $p(\textbf{s},j,\mathbf{J})$:
\begin{equation} \label{eq:probability}
{\cal P}^{\mathcal{S}} (\textbf{s};i,j;\mathbf{J}) =
\frac{p(\textbf{s},j,\mathbf{J})}{\displaystyle \sum_{k \in
{\mathcal{N}_i}} p(\textbf{s},k,\mathbf{J})}.
\end{equation}
We can possibly restrict the choice just to the set
$\tilde{\mathcal{N}}_i$, hence defining $\tilde{{\cal
P}}^{\mathcal{S}} (\textbf{s};i,j;\mathbf{J}) =
p(\textbf{s},j,\mathbf{J})/ \sum_{k \in {\tilde{\mathcal{N}}_i}}
p(\textbf{s},k,\mathbf{J})$. Notice that, as we have verified, the
use of $\tilde{{\cal P}}^{\mathcal{S}}$ instead of ${\cal
P}^{\mathcal{S}}$ does not imply any qualitative change in the
results.

- According to the normalized probability ${\cal P}^{\mathcal{S}}$
(see Eq.~\ref{eq:probability}), we extract randomly the node
$\bar{i} \in \mathcal{N}_i$ and realize the opinion flip
$s_{\bar{i}} \rightarrow - s_{\bar{i}}$.

- We set $\bar{i} \equiv i$ and we iterate the procedure.

Finally, it should be underlined that in this dynamics detailed balance is explicitly violated \cite{BBCV2002,ABCV2005}; indeed, its purpose is not to recover a canonical Bolzmann equilibrium but rather to model possible mechanism making the system evolve,
and ultimately, to describe, at an effective level, the statics
reached by a ``socially plausible'' dynamics for opinion
spreading \cite{prima}.

\subsection{Equilibrium behavior}

The diffusive dynamics was shown to be able to lead the system
toward a well defined steady state and to recover the expected phase
transition, although the critical temperature revealed was larger
than the expected one \cite{BBCV2002}. Such results were also shown to be robust with respect to the the spin magnitude \cite{ABCV2005} and the underlying topology \cite{prima}.

\begin{figure}[tb]
\includegraphics[height=60mm]{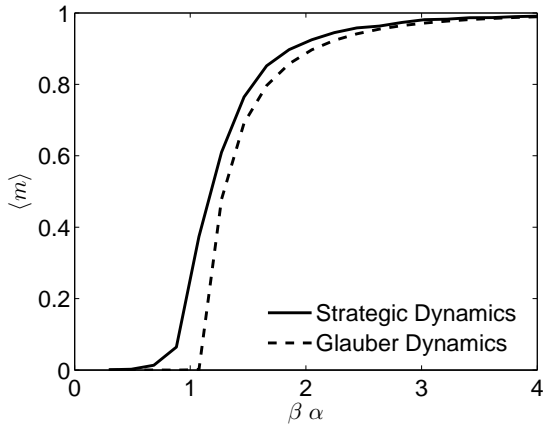}
\caption{\label{fig:p3} Critical behavior of the magnetization for
the two dynamics (diffusive and standard Glauber) as a function of
$\beta$ and fixed $\alpha=10$. The former dynamics gives rise to a critical point
higher with respect to the latter.} 
\end{figure}

In particular, it was evidenced that there exists a critical value of the
parameter $\beta_c^{\mathcal{S}}$ below which the system is spontaneously
ordered. However, $\beta_c^{\mathcal{S}}$ was found to be
appreciably smaller than the critical value $\beta_c(\alpha)$
expected for the canonical Ising model on a Erdos-Renyi random graph.
Interestingly, it is not possible describe the system subjected to the diffusive dynamics by introducing an effective Hamiltonian obtained  from Eq (\ref{IsingER}) by a trivial rescaling.
In fact, we consider the dependence on the magnetization displayed by the energies $E^{\mathcal{S}}(m)$ and $E(m)$, measured for system evolving according to the diffusive dynamics and to a traditional dynamics, respectively. As for the latter, from Eq. (\ref{IsingER}) it is easy to see that $E = m^2$. As for $E^{\mathcal{S}}(m)$, we found that $E^{\mathcal{S}}<E$ for $0<m<1$, while $E^{\mathcal{S}}=E$ for $m=0$ and $m=1$. This is compatible with a power law behaviour $E^{\mathcal{S}} \sim m^ {2+\epsilon}$. In order to obtain an estimate for $\epsilon$ we measured the ratio $E^{\mathcal{S}}/E$ as a function of $m$; data are shown in the log-log scale plot of Fig.~2. Now, fitting procedures suggest that $\epsilon \approx 0.15$. Notice that the large deviations from the linear behaviour evidenced at small value of $\langle m \rangle$ are due to the fact that we are dividing two quantities close to zero.

So our idea is the following: as the critical temperature raises with $p$ ranging
from two to infinity, there can be a suitable real value of $p$
that matches the critical temperature found numerically, and this is compatible with the plots of $E^{\mathcal{S}}(m)$ and $E(m)$.
Notice that  for $p>2, p\in\mathbb{N}$, ferromagnetic transitions
are no longer critical phenomena. At the critical line ``jumps'' in the magnetization and a latent heat do exist. However, if $p$ is
though of as real, for $p$ slightly bigger than two, as suggested by our data,
such a jump should be small (and it goes to zero whenever $p\to2$), so it is
difficult to check it just by looking at the magnetization as a
function of $\alpha,\beta$. To investigate this property, noting
that the discontinuity of the entropy (latent heat) can be
checked by looking at the compression rate of the phase space via
the Shannon Theorem, we show the compression
rate of the strategic dynamics $R_s$ normalized by the compression rate
of the Glauber one $R$, which offers another indication of the
presence of a $p>2$ behavior. In fig.~2, bottom panel, we plot $C=R_s/R$ as a function of the inverse temperature $\beta$; a minimum occurs just at the critical temperature of the system evolving according to the strategic dynamics.

\begin{figure}[tb]\bc
\includegraphics[width=70mm,height=50mm]{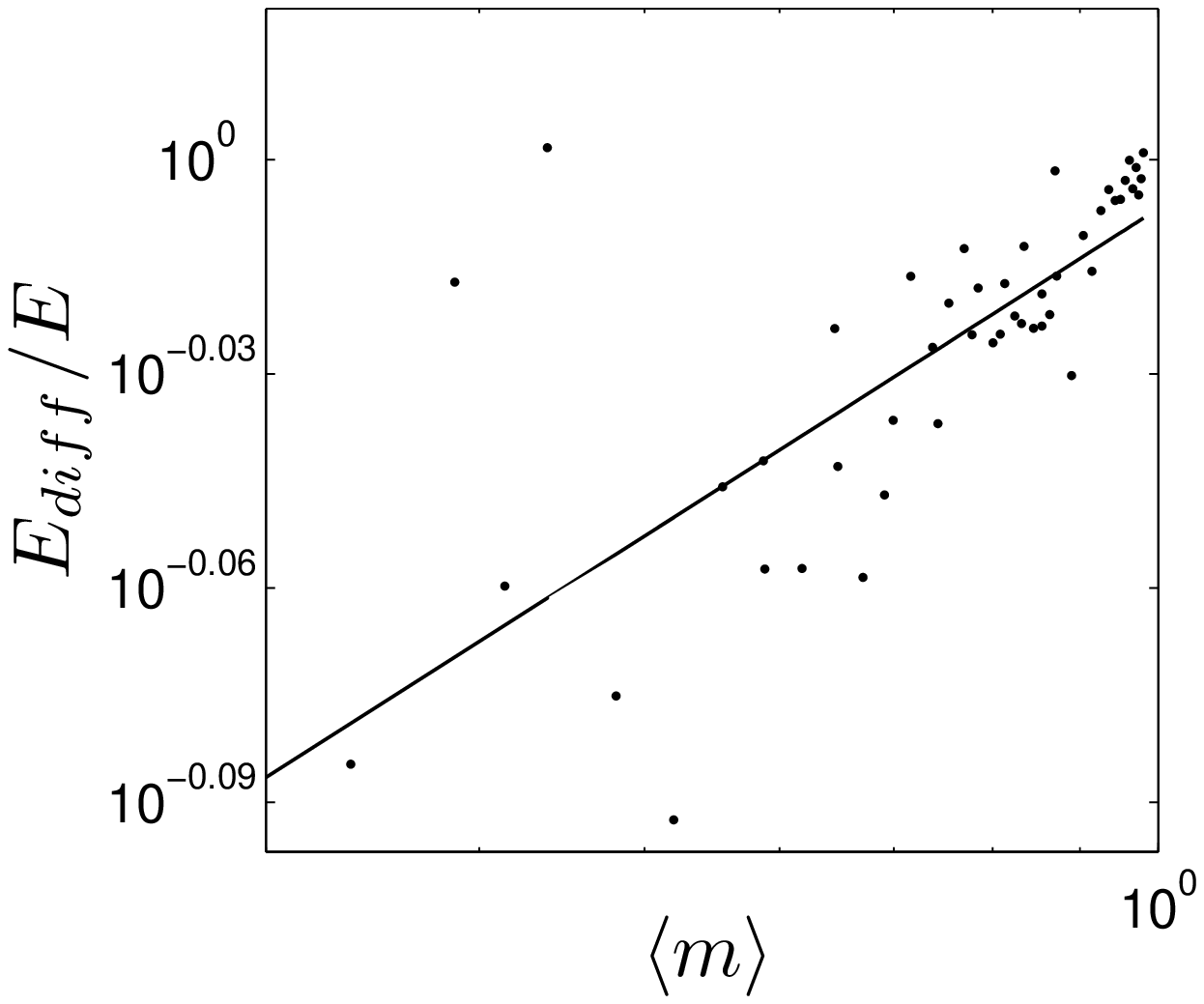}
\includegraphics[width=60mm,height=50mm]{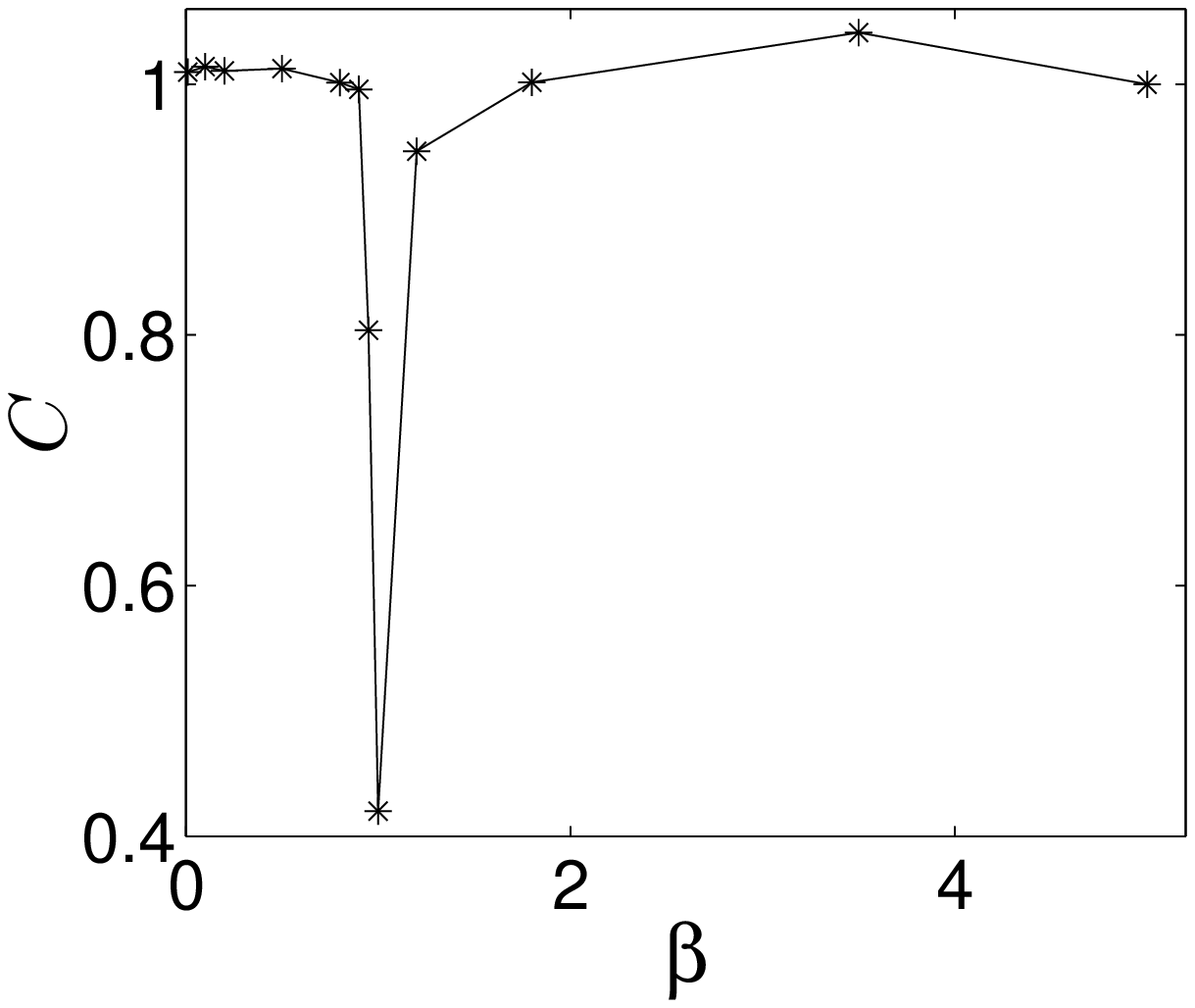}
\caption{\label{fig:p3}Up: We extrapolate from $ m(\beta)$ and $E (\beta)$ the
plot $E(m)$ for both the dynamics. It is worth noting that the
diffusive dynamics deals to a curve living always ''below" the one
obtained by Glauber dynamics results. This strongly suggest the
cooperation of $p>2$ spins each interaction. Down: We show the phase space compression
ratio $C=R_s/R$, where $R_s$ is the phase space compression
of the strategic dynamics and $R$ stands for the standard Glauber
one.}\ec
\end{figure}

\section{Statics of many body interactions}\label{tre}

In this section we introduce a  $p$-spin model to match the steady
state recovered by the diffusive dynamics.

\subsection{The diluted even mean field  $p$-spin ferromagnet}

In this section we exploit the properties of a diluted even
$p$-spin ferromagnet: we restrict ourselves only to even values of
$p$ for mathematical convenience (the investigation with the
cavities is much simpler), but, due to monotonicity of all the
observables among $p$, there is no need to think at this as a real
restriction (furthermore simulations on odd values of $p$ confirm
this statement).
\newline
First of all, we define a suitable Hamiltonian acting on a
Erdos-Renyi random graph, with connectivity $\alpha$, made up by
$N$ agents $\sigma_{i}=\pm1, \ i \in [1,N]$.
\newline
Introducing $p$ families
$\{i_{\nu}^1\},\{i_{\nu}^2\},...,\{i_{\nu}^p\}$ of i.i.d. random
variables uniformly distributed on the previous interval, the
Hamiltonian is given by the following expression
\begin{equation}\label{ham}
H_{N}(\sigma,\gamma(\alpha))=-\sum_{\nu=1}^{k_{\gamma (\alpha) N}}
\sigma_{i_{\nu}^1}\sigma_{i_{\nu}^2}...\sigma_{i_{\nu}^p}
\end{equation}
where, reflecting the underlying network, $k$ is a Poisson
distributed random variable with mean value $\gamma(\alpha) N$.
The relation among the coordination number $\alpha$ and $\gamma$
is $\gamma \propto \alpha^{p-1}$: this will be easily understood a
few line later by a normalization argument coupled with the high
connectivity limit of this mean field model.

The quenched expectation of the model is given by the composition
of the Poissonian average with the uniform one performed over the
families $\{i_{\nu}\}$
\begin{equation}
\textbf{E}[\cdot] = E_PE_i[\cdot] = \sum_{k=0}^{\infty}
\frac{e^{-\gamma(\alpha) N}(\gamma(\alpha)
N)^k}{k!N^p}\sum_{i_{\nu}^1....i_{\nu}^p}^{1,N}[\cdot].
\end{equation}

As they will be useful in our derivation, it is worth stressing
the following properties of the Poisson distribution: Let us
consider a function $g:\mathbb{N}\to\mathbb{R}$, and a Poisson
variable $k$ with mean $\gamma N$, whose expectation is denoted by
$\mathbb{E}$.
\newline
It is easy to verify that
\begin{eqnarray}\label{Pp1}
\mathbb{E}[k g(k)] &=& \gamma N \mathbb{E} [g(k-1)] \\ \label{Pp2}
\partial_{\gamma N}\mathbb{E}[g(k)] &=&
\mathbb{E}[g(k+1)-g(k)]\\ \label{Pp3}
\partial^2_{(\gamma N)^2}\mathbb{E}[g(k)] &=&
\mathbb{E}[g(k+2)-2g(k+1)+g(k)].
\end{eqnarray}

The Hamiltonian (\ref{ham}), as written, has the advantage that it
is the sum of (a random number of) i.i.d.\ terms. To see the
connection to a more familiar Hamiltonian wrote in terms of
adjacency tensor $A_{i_1,...,i_p}$, we note that the
Poisson-distributed total number of bonds obeys $P_{\gamma
N}=\gamma N + O(\sqrt{N})$ for large $N$. As there are $N^p$
ordered spin p-plets $(i_1,...,i_p)$, each gets a bond with
probability $\sim \alpha/N$ for large $N$. The probabilities of
getting two, three (and so on) bonds scale as $1/N^2,1/N^3,\ldots$
so can be neglected. The probability of having a bond between any
unordered p-plet of spins is $p!$ as large, i.e.\ $2\alpha/N$ for
$p=2$.
\newline
It is possible to show that our version of the Hamiltonian in fact
is thermodynamically equivalent with the more familiar involving
the explicit adjacency tensor $A_{i_1,...,i_p}$, by recall at
first both the models
\begin{equation}\label{parag}
-H_N(\sigma; k) \sim - \hat{H}_N(\sigma) = \sum_{i_1,...,i_p}^N
A_{i_1,...,i_p}\sigma_{i_1}...\sigma_{i_p},
\end{equation}
where $k$ is a Poisson variable with mean $\gamma N \sim
\alpha^{p-1} N$ and $A_{i_1,...,i_k}$ are all independent Poisson
variables of mean $\gamma/N^{p-1} \sim (\alpha / N)^{p-1}$.

Then, it is enough to consider the streaming of  the following
interpolating free energy (whose structure proves the statement a
priori by its thermodynamics meaning), depending on the real
parameter $t\in[0,1]$
$$
\phi(t) = \frac{\mathbb{E}}{N}\ln\sum_{\sigma}e^{\beta
\large(\sum_{\nu=1}^k \sigma_{i^1_{\nu}}...\sigma_{i^p_{\nu}} +
\sum_{i_1,...,i_p}^N A_{i_1,...,i_p}
\sigma_{i_1}...\sigma_{i_p}\large)},
$$
where $k$ is a Poisson random variable with mean $\gamma N t$ and
$A_{i_1,...,i_p}$ are random Poisson  variables of mean
$(1-t)\gamma/N^{p-1}$, so note that the two models separated are
recovered in the two extremals of the interpolation (for $t=0,1$).
By computing the $t$-derivative, we get
\begin{eqnarray}
\frac{1}{\gamma}\frac{d \phi(t)}{dt} &=& \mathbb{E}\ln(1+
\Omega(\sigma_{i_0^1}...\sigma_{i_0^p})\tanh(\beta))
\\ \nonumber &-& \frac{1}{N^{p}}\sum_{i_1,...,i_p}^N\ln(1+\Omega(\sigma_{i_1}...\sigma_{i_p})\tanh(\beta))=0,
\end{eqnarray}
where the label $0$ in $i_0^k$ stands for a new spin, born in the
derivative, accordingly to the Poisson property (\ref{Pp2}); as
the $i_0$'s are independent of the random site indices in the
$t$-dependent $\Omega$ measure, the equivalence is proved.

\bigskip

Following a statistical mechanics approach, we know that the
macroscopic behavior, versus the connectivity $\alpha$ and the
inverse temperature $\beta$, is described by the following free energy
density
\begin{eqnarray}
A(\alpha,\beta) &=& \lim_{N \to \infty} A_N(\alpha,\beta) \\
\nonumber &=& \lim_{N \to
\infty}\frac1N\textbf{E}\ln\sum_{\sigma}\exp(-\beta
H_{N}(\sigma,\gamma(\alpha))).\
\end{eqnarray}
The normalization constant can be checked by performing  the
expectation value of the cost function:
\begin{eqnarray}\nonumber
\textbf{E}[H] &=& -\gamma N m^p \\  \textbf{E}[H^2] -
\textbf{E}^2[H] &=& \gamma^2 N^2\Big[(q_{12}^p - m^p\Big) +
O(\frac{1}{N})\Big],
\end{eqnarray}
by which it is easy to see that the model is well defined, in
particular it is linearly extensive in the volume. Then, in the
high connectivity limit each agent interacts with all the others
($\alpha \sim N$) and, in the thermodynamic limit, $\alpha \to
\infty$. Now, if $p=2$ the amount of couples in the summation
scales as $N(N-1)/2$ and, with $\gamma = 2\alpha$, a linear
divergence of $\alpha$ (desired to get a finite ratio $\alpha/N$
for each coupling) provides the right scaling; if $p=3$ the amount
of triples scales as $N(N-1)(N-2)/3!$ and, with $\gamma =
3!\alpha^2$, again we find the right connectivity behavior. The
generalization to every finite $p<N$ is straightforward.

\subsection{Properties of the random diluted  $p$-spin model}

Let us now introduce the whole statistical mechanics machinery: we
start by  the partition function defined as
\begin{equation}
Z_N(\gamma,\beta) = \sum_{\{\sigma\}}e^{-\beta
H_N(\sigma,\gamma)},
\end{equation}
the quenched pressure can be written as
\begin{equation}\nonumber
A_{N}(\gamma,\beta) = \frac{1}{N}\textbf{E}\ln Z_N(\gamma,\beta),\
\end{equation}
the Boltzmann state is given by
\begin{equation}
\omega(g(\sigma)) = \frac{1}{Z_{N}(\gamma,\beta)}
\sum_{\{\sigma_N\}} g(\sigma) e^{-\beta H_{N}(\sigma,\gamma)},
\end{equation}
with its replicated form
\begin{equation}
\Omega(g(\sigma)) = \prod_s \omega^{(s)}(g(\sigma^{(s)}))
\end{equation}
and the total average $\langle g(\sigma)\rangle$ is defined as
\begin{equation}
\langle g(\sigma)\rangle = \textbf{E}[\Omega(g(\sigma))].
\end{equation}
\medskip
\newline
Let us introduce further, as order parameters of the theory, the
multi-overlaps \be
q_{1...n}=\frac1N\sum_{i=1}^{N}\sigma^{(1)}_{i}...\sigma^{(n)}_{i},
\ee with a particular attention at the magnetization $m = q_1
=(1/N)\sum_{i=1}^{N}\sigma_{i}$ and to the two replica overlap
$q_{12} = (1/N)\sum_{i=1}^{N}\sigma_{i}^1\sigma_i^2$.

Before starting our free energy analysis, we want to point out
also  the connection among this diluted version and the fully
connected counterpart.
\newline
Let us remember that the Hamiltonian of the fully connected  $p$-spin
model (FC) can be written as \cite{abarra3} \be H^{FC}_{N}(\sigma)
= \frac{p!}{2 N^{p-1}}\sum_{1 \leq i_1 < ... < i_p \leq N}
\sigma_{i_1}\sigma_{i_2}...\sigma_{i_p}, \ee and let us consider
the  trial function $\hat{A}(t)$ defined as follows \be \hat{A}(t)
= \frac{1}{N}\mathbb{E}\ln \sum_{\sigma} \exp\Big[ \beta
\sum_{\nu}^{P_{\gamma N
t}}\sigma_{i_{\nu}^1}\sigma_{i_{\nu}^2}...\sigma_{i_{\nu}^p} +
(1-t)\frac{\beta' N}{2}m^p \Big], \ee which interpolates between
the fully connected  $p$-spin model and the diluted one, such that
for $t=0$ only the fully connected survives, while the opposite
happens for $t=1$. Let us work out the derivative with respect to
$t$ to obtain \begin{eqnarray} \partial_t \hat{A}(t) &=&
(p-1)\alpha^{p-1}\ln\cosh(\beta) \\ \nonumber &-& (p-1)
\alpha^{p-1}\sum_{n}\frac{-1^n}{n}\theta^n \langle q_n^p\rangle -
\frac{\beta'}{2}\langle m^p \rangle, \end{eqnarray} by which we
see that the correct scaling, in order to recover the proper
infinite connectivity model, is obtained when $\alpha \to \infty$,
$\beta \to 0$ and $\beta' = 2(p-1) \alpha^{p-1}\tanh(\beta)$ is
held constant.

\begin{remark}
It is worth noting that in social modeling, usually, the role of
the temperature is left, or at least coupled together, to the
interaction strength $J$. As a consequence, in order to keep
$\beta'$ fixed, on different network dilution, the strength must
be rescaled accordingly to
$$
J = \tanh^{-1}\Big( \frac{\beta'}{2(p-1)\alpha^{p-1}} \Big),
$$
while, if present, an external field remains unchanged as it is a
one-body term, like $h\sum_i^N  \sigma_i$, unaffected by dilution.
\end{remark}
\begin{remark}
The dilute  $p$-spin model reduces to the fully connected one, in the
infinite connectivity limit, uniformly in the size of the system.
\end{remark}

\subsection{The smooth cavity approach}

In this section we want to look for an iterative expression of the
free energy density by using a version of the cavity strategy
\cite{barra1, abarra} that we briefly recall: the idea
behind the cavity techniques \cite{guerra2,MPV}, which, for
our purposes, resembles the stochastic stability
approach \cite{ac,parisiSS}, is that information concerning
the free energy density can be extrapolated when looking at the
incremental extensive free energy given by the addition of a spin.
\newline
In diluted models, this pasted spin changes also (infinitesimally
in the high $N$ limit) the connectivity and in evaluating how the
free energy density varies conformingly with this, we are going to
prove that it can be written in terms of a cavity function and
such a connectivity shift.
\newline
So the behavior of the system is encoded into these two terms. The
latter is simpler as it is made only by stochastically stable
terms (a proper definition of these terms will follow through the
section). The former, instead, must be expressed via these terms
and this will be achieved by iterative expansions.
\newline
At first we show how the free energy density can be decomposed via
these two terms (the cavity function and the connectivity shift).
Then, we analyze each term separately. We will see that they
can be expressed by the momenta of the magnetization and of the
multi-overlaps, weighted in a perturbed Boltzmann state, which
recovers the standard one in the thermodynamic limit.
\begin{theorem}\label{primolevi}
\textit{In the thermodynamic limit, the quenched pressure of the
even  $p$-spin diluted ferromagnetic model is given by the
following expression} \be A(\alpha,\beta) = \ln2
-\frac{\alpha}{p-1}\frac{d}{d\alpha}A(\alpha,\beta) +
\Psi(\alpha,\beta,t=1), \ee
\end{theorem}
where the cavity function $\Psi(t,\alpha,\beta)$ is introduced as
\begin{eqnarray}\label{cavity1}
&& \textbf{E}\Big[\ln\frac{\sum_{\{\sigma\}}
e^{\beta\sum_{\nu=1}^{k_{\tilde{\gamma}N}}
\sigma_{i_{\nu}^1}\sigma_{i_{\nu}^2}...\sigma_{i_{\nu}^p}}\;
e^{\beta\sum_{\nu=1}^{k_{2\tilde{\gamma}t}}
\sigma_{i_{\nu}^1}\sigma_{i_{\nu}^2}...\sigma_{i_{\nu}^{p-1}}}}
{\sum_{\{\sigma\}} e^{\beta\sum_{\nu=1}^{k_{\tilde{\gamma}N}}
\sigma_{i_{\nu}^1}\sigma_{i_{\nu}^2}...\sigma_{i_{\nu}^p}}}\Big]= \nonumber \\
&&  \textbf{E}\Big[\ln \frac{Z_{N,t}(\tilde{\gamma},\beta)}
{Z_{N}(\tilde{\gamma},\beta)}\Big] =
\Psi_N(\tilde{\gamma},\beta,t),
\end{eqnarray}
with \be\label{cavity2} \Psi(\gamma,\beta,t) =
\lim_{N\rightarrow\infty}\Psi_N(\tilde{\gamma},\beta,t). \ee
\medskip
\newline
For the sake of clearness, to avoid interrupting the paper with a
long technical calculation, the proof of the Theorem is reported
in the Appendix.
\newline
Thanks to the previous theorem, it is possible to figure out an
expression for the pressure by studying the properties of the
cavity function $\Psi(\alpha,\beta)$ and the connectivity shift
$\partial_{\alpha}A(\alpha,\beta)$.
\newline
Using the properties of the Poisson distribution (\ref{Pp1},
\ref{Pp2}), we can write
\begin{eqnarray}
\frac{d}{d\alpha}A(\alpha,\beta)
&=& \frac{(p-1)}{N}\alpha^{p-2}\frac{d}{d\gamma}\textbf{E}\Big[\ln Z_N(\gamma,\beta)\Big] = \nonumber \\
&=& (p-1)\alpha^{p-2}\textbf{E}\Big[\ln \sum_{\{\sigma\}}
e^{\beta\sum_{\nu=1}^{k+1}
\sigma_{i_{\nu}^1}...\sigma_{i_{\nu}^p}} -  \nonumber \\ \nonumber
&-& \ln \sum_{\{\sigma\}} e^{\beta\sum_{\nu=1}^{k}
\sigma_{i_{\nu}^1}...\sigma_{i_{\nu}^p}}\Big]. \nonumber
\end{eqnarray}
Now considering the relation (and definition)
\begin{eqnarray}
e^{\beta\sigma_{i_0^1}...\sigma_{i_0^p}} &=&
\cosh\beta + \sigma_{i_0^1}...\sigma_{i_0^p}\sinh\beta, \\
\theta &=& \tanh\beta,
\end{eqnarray}
we can write
\begin{eqnarray}
&& \frac{d}{d\alpha}A(\alpha,\beta) = \\ \nonumber &&
(p-1)\alpha^{p-2}\Big[\ln\cosh\beta + \textbf{E}[\ln(1 +
\omega(\sigma_{i_{\nu}^1}...\sigma_{i_{\nu}^p})\theta)]\Big].
\end{eqnarray}
At the end, expanding the logarithm, we obtain
\begin{eqnarray}
\frac{d}{d\alpha}A(\alpha,\beta) &=&
(p-1)\alpha^{p-2}\ln\cosh\beta - \\ \nonumber
 &-& (p-1)\alpha^{p-2}\sum_{n=1}^{\infty}\frac{(-1)^n}{n}\theta^n
\langle q_{1,...,n}^p \rangle.
\end{eqnarray}
\medskip
\newline
With the same procedure it is possible to show that
\begin{eqnarray}\label{Psit}
\frac{d}{dt}\Psi(\tilde{\alpha},\beta,t) &=&
2\tilde{\alpha}^{p-1}\ln\cosh\beta - \\ \nonumber &-&
2\tilde{\alpha}^{p-1}\sum_{n=1}^{\infty}\frac{(-1)^n}{n}\theta^n
\langle q_{1,...,n}^{p-1} \rangle_{\tilde{\alpha},t},
\end{eqnarray}
by which,we see that even the cavity function, once integrated
back against $t$ the r.h.s. of eq.(\ref{Psit}),  can be expressed
via all the order parameters of the model.

\begin{equation}\nonumber
\Psi(\tilde{\alpha},\beta,t)=2\tilde{\alpha}^{p-1}\Big(\ln\cosh(\beta)-
\sum_{n=1}^{\infty}\frac{(-\theta)^n}{n}\int_0^t \langle
q_{1,...,n}^{p-1}\rangle_{\tilde{\alpha},t} \Big).
\end{equation}

So we can understand the properties of the free energy by analyzing
the properties of the order parameters: magnetization and
overlaps, weighted in their extended Boltzmann state
$\tilde{\omega}_t$.
\newline
Further, as we expect the order parameters being able to describe
thermodynamics even in the true Boltzmann states $\omega, \Omega$
\cite{landau}, accordingly to the following definitions, we are
going to show that {\em filled} order parameters (the ones
involving even numbers of replicas) are stochastically stable, or
in other words, are independent by the $t$-perturbation in the
thermodynamic limit, while the others, not filled, become filled,
again in this limit (such that for them $\omega_t\to \omega$ in
the high $N$ limit and thermodynamics is recovered). The whole is
explained in the following definitions and theorems of this
section.

\begin{definition}
We define the t-dependent Boltzmann state $\tilde{\omega}_t$ as
\begin{eqnarray}\label{dente}
&& \tilde{\omega}_t(g(\sigma)) = \\ \nonumber &&
\frac{1}{Z_{N,t}(\gamma,\beta)} \sum_{\{\sigma\}}g(\sigma)
e^{\beta\sum_{\nu=1}^{k_{\tilde{\gamma}N}}
\sigma_{i_{\nu}^1}...\sigma_{i_{\nu}^p} +
\beta\sum_{\nu=1}^{k_{2\tilde{\gamma}t}}
\sigma_{i_{\nu}^1}...\sigma_{i_{\nu}^{p-1}}}\label{dante},
\end{eqnarray}
where $Z_{N,t}(\gamma,\beta)$ extends the classical partition
function in the same spirit of the numerator of eq.(\ref{dente})
itself.
\end{definition}
We see that the original Boltzmann state of a $N$-spin system is
recovered by sending $t \to 0$, while, sending $t \to 1$ and
gauging the spins, it is possible to build a Boltzmann state of a
$N+1$ spins, with a little shift both in $\alpha,\beta$, which
vanishes in the $N \to \infty$ limit, as prescribed in $(22,23)$.
\newline
Coherently with the implication of thermodynamic limit (by which
$A_{N+1}(\alpha,\beta)-A_N(\alpha,\beta) =0$ for $N \to \infty$),
we are going to define the {\em filled} overlap monomials and show
their independence (stochastic stability) with respect to the
perturbation encoded by the interpolating parameter $t$.
\newline
These parameters are already ''good" order parameters describing
the theory, while the others (the not-filled ones) must be
expressed via the formers, and this will be achieved by expanding
them with a suitably introduced streaming equation.

\begin{definition} We can split the class of monomials of the
order parameters in two families:
\begin{itemize}

\item We define {\itshape filled} or equivalently {\itshape stochastically
stable} those overlap monomials with all the replicas appearing an
even number of times (i.e. $q_{12}^2$,\ $m^2$,\
$q_{12}q_{34}q_{1234}$).

\item We define {\itshape non-filled}
those overlap monomials with at least one replica appearing an odd
number of times (i.e. $q_{12}$,\ $m$,\ $q_{12}q_{34}$).
\end{itemize}
\end{definition}
We are going to show three theorems that will play a guiding role
for our expansion: as this approach has been deeply developed in
similar contexts (as fully connected Ising and  $p$-spin models
\cite{abarra3,abarra}, fully connected spin glasses \cite{barra1}
or diluted ferromagnetic models \cite{ABC,BCC}, which are the
''boundaries'' of the model of this paper) we will not show all
the details of the proof, but we sketch them in the appendix as
they are really intuitive. The interested reader will found a
clear derivation in the appendix but can deepen this point by
looking at the original works.
\begin{theorem}\label{ciccia}
In the thermodynamic limit and setting $t=1$ we have \be
\tilde{\omega}_{N,t}(\sigma_{i_1}\sigma_{i_2}...\sigma_{i_n}) =
\tilde{\omega}_{N+1}(\sigma_{i_1}\sigma_{i_2}...\sigma_{i_n}\sigma_{N+1}^n).
\ee
\end{theorem}
\begin{theorem}\label{saturabili}
Let $Q_{ab}$ be a not-filled monomial of the overlaps (this means
that $q_{ab}Q_{ab}$ is filled). We have \be
\lim_{N\rightarrow\infty}\lim_{t\rightarrow1} \langle Q_{ab}
\rangle_t = \langle q_{ab}Q_{ab} \rangle, \ee (examples:
\newline for $N \rightarrow \infty$ we get $\langle m_1 \rangle_t
\rightarrow \langle m_1^2 \rangle,\quad \langle q_{12} \rangle_t
\rightarrow \langle q_{12}^2 \rangle$).
\end{theorem}
\begin{theorem}\label{saturi}
In the $N\rightarrow\infty$ limit, the averages
$\langle\cdot\rangle$ of the filled polynomials are t-independent
in $\beta$ average.
\end{theorem}
\medskip

\subsection{Properties of the free energy}

In this section we are going to address various points: at first
we work out the constraints that the model must fulfil, which are
in agreement both with a self-averaging behavior of the
magnetization as with the replica-symmetric behavior of the
multi-overlaps \cite{gulielmo}; then we write an iterative
expression for the free energy density and its links with known
models as diluted ferromagnets ($p\to2$ limit) and fully connected
$p$-spin models ($\alpha \to \infty$ limit).

\bigskip

With the following definition \begin{eqnarray} \tilde{\beta} &=&
2(p-1)\tilde{\alpha}^{p-1}\theta \\ \nonumber &=&
2(p-1)\alpha^{p-1}\frac{N}{N+1}\theta \quad
\stackrel{N\rightarrow\infty}{\longrightarrow}
2(p-1)\alpha^{p-1}\theta = \beta', \end{eqnarray} we show (and
prove in the appendix) the streaming of replica functions, by
which not filled multi-overlaps can be expressed via filled ones.
\begin{proposition}\label{stream}
Let $F_s$ be a function of s replicas. Then the following
streaming equation holds
\begin{eqnarray}
\frac{\partial\langle F_s \rangle_{t,\tilde{\alpha}}}{\partial t}
&=& \tilde{\beta} \Big[\sum_{a=1}^s\langle F_s
m_a^{p-1}\rangle_{t,\tilde{\alpha}} - s \langle F_s
m_{s+1}^{p-1}\rangle_{t,\tilde{\alpha}}\Big] \quad
\\ \nonumber
&+& \tilde{\beta}\theta \Big[ \sum_{a<b}^{1,s}\langle F_s
q_{a,b}^{p-1} \rangle_{t,\tilde{\alpha}} - s\sum_{a=1}^s\langle
F_s q_{a,s+1}^{p-1}\rangle_{t,\tilde{\alpha}} \\ \nonumber &+&
\frac{s(s+1)}{2!}\langle F_s
q_{s+1,s+2}^{p-1}\rangle_{t,\tilde{\alpha}}\Big] + O(\theta^3).
\end{eqnarray}
\end{proposition}
\begin{remark}
We stress that, at the first two level of approximation presented
here, the streaming has has the structure of a $\theta$-weighted
linear sum of the Curie-Weiss streaming ($\theta^0$ term)
\cite{abarra} and the Sherrington-Kirkpatrick streaming
($\theta^1$ term) \cite{barra1}, conferring a certain degree of
independence by the kind of quenched noise (frustration or
dilution) to mathematical structures of disordered systems.
\end{remark}

\bigskip

It is now immediate to obtain the linear order parameter
constraints (often known as Aizenman-Contucci polynomials
\cite{abarra,ac,BCC}) of the theory: in fact, the generator of
such a constraint is the streaming equation when applied on each
filled overlap monomial (or equivalently it is possible to apply
the streaming on a not-filled one and then gauge the obtained
expression; for the sake of clearness both the methods will be
exploited, the former for $q_2$ and the latter for $m$).
\newline
As examples, dealing with the terms $m^{p-1}$ and $q_{2}^{p-1}$,
it is straightforward to check that \begin{eqnarray}\nonumber 0
&=& \lim_{N \to \infty} \frac{\partial\langle m_N^{p-1}
\rangle_{t,\tilde{\alpha}}}{\partial t} = \tilde{\beta} \Big(
\langle m_1^{2(p-1)} \rangle - \langle m_1^{p-1} \rangle^2 \Big)\\
\nonumber &+& \tilde{\beta}\theta \Big( \langle m_1^{p-1}q_2^{p-1}
\rangle - \langle m_1^{p-1}\rangle \langle  q_2^{p-1} \rangle
\Big) + O(\theta^3), \nonumber \end{eqnarray} then, by gauging the
above expression, in the thermodynamic limit, (as $\lim_{N \to
\infty}\langle m_N^{p-1}\rangle_t \to \langle m^p \rangle$), we
get
$$
\Big( (\langle m_1^{2p} \rangle - \langle m_1^p \rangle^2) +
\theta ( \langle q_{2}^{2p} \rangle - \langle q_{2}^p \rangle^2
)\Big)=0 \ \ \forall \theta \in \mathbb{\mathcal{R}}^+,
$$
which, as holds for every $\theta$ suggests both self-averaging
for the energy (by which all the linear constraints can be
derived\cite{BCC}) due to the first term, as well as replica
symmetric behavior of the two replica overlap due to the last one.
\newline
In the same way, the contribution of the $\langle q_{2}^2\rangle$
generator is  \begin{eqnarray}  \nonumber 0 &=& \Big( (\langle
q_{12}^{p-1} m_1^{p-1} \rangle + \langle q_{12}^{p-1} m_2^{p-1}
\rangle - 2 \langle q_{12}^{p-1} m_3^{p-1} \rangle) +  \\
\nonumber &+& \theta ( \langle q_{12}^{p-1} q_{12}^{p-1} \rangle -
4 \langle q_{12}^{p-1} q_{23}^{p-1} \rangle + 3 \langle
q_{12}^{p-1} q_{34}^{p-1} \rangle) \Big), \end{eqnarray} which
shows replica symmetric behavior of the magnetization by the first
term and the classical Aizenman-Contucci relations
\cite{ac,BCC} by the latter.

\bigskip

Furthermore, turning now our attention to the free energy, it is
easy to see that the streaming equation  allows to generate all
the desired overlap functions coupled to every well behaved $F_s$.
In this way, if $F_s$ is a not filled overlap, we can always
expand recursively it into a filled one, with the only price to
pay given by the $\theta$ order that has to be reached or, which
is equivalent, the number of derivatives that have to be
performed.
\newline
Let us now remember the t-derivative of the cavity function
(\ref{Psit}), showing explicitly the first two terms of its
expansion
\begin{eqnarray}\label{Psit2}
\frac{d}{dt}\Psi(\tilde{\alpha},\beta,t) &=&
2\tilde{\alpha}^{p-1}\ln\cosh\beta + \tilde{\beta}\langle
m_1^{p-1} \rangle_{\tilde{\alpha},t} - \\ \nonumber  &-&
\frac{\tilde{\beta}}{2}\theta\langle q_{12}^{p-1}
\rangle_{\tilde{\alpha},t} -
2\tilde{\beta}^{p-1}\sum_{n=3}^{\infty}\frac{-1^n\theta^n}{n}
\langle q_{1,...,n}^{p-1} \rangle_{\tilde{\alpha},t}.
\end{eqnarray}

As derivative of fillable terms involve filled ones, we can arrive
to an analytical form of $\Psi(\alpha,\beta)$ if we calculate it
as the t-integral of its t-derivative, together with the obvious
relation $\Psi(t=0) = 0$. So, if we apply the streaming equation
machinery to the overlaps constituting equation (\ref{Psit2}), we
are able to fill them and to get them free from the t-dependence
in the thermodynamic limit. In this way we are allowed to bring
them out from the final t-integral.
\newline
In fact, without gauging (so, not only in the ergodic regime,
where symmetries are preserved), we can expand the streaming of
$\langle m^{p-1}\rangle_t$:
\newline
\begin{eqnarray}
\frac{d\langle m_1^{p-1} \rangle_t}{dt} &=&
\tilde{\beta}\Big[\langle m_1^{2(p-1)} \rangle - \langle m_1^{p-1}m_2^{p-1} \rangle_t\Big] + \nonumber \\
&-&  \tilde{\beta}\theta\Big[\langle m_1^{p-1}q_{12}^{p-1}
\rangle_t - \langle m_1^{p-1}q_{23}^{p-1} \rangle_t\Big] +
O(\theta^2). \nonumber
\end{eqnarray}
We can note the presence of the filled monomial $\langle
m_1^{2(p-1)} \rangle$, whose t-dependence has been omitted
explicitly to underly its stochastic stability, while the overlaps
$\langle m_1^{p-1}m_2^{p-1} \rangle_t$ and $\langle
m_1^{p-1}q_{12}^{p-1} \rangle_t$ can be saturated in two steps of
streaming. This will be sufficient, wishing to have a fourth order
expansion for the cavity function.
\newline
We now derive these two functions and apply the same scheme to all
the overlaps that appear and that have to be necessary filled in
order to obtain the desired result.
\begin{eqnarray}
&& \frac{d\langle m_1^{p-1}m_2^{p-1} \rangle_t}{dt} = \nonumber \\
&& 2\tilde{\beta}\Big[
\langle m_1^{2(p-1)}m_2^{p-1} \rangle_t - \langle m_1^{p-1}m_2^{p-1}m_3^{p-1} \rangle_t\Big] + \nonumber \\
&& \theta \tilde{\beta} \Big[\langle
m_1^{p-1}m_2^{p-1}q_{12}^{p-1} \rangle -
4\langle m_1^{p-1}m_2^{p-1}q_{13}^{p-1} \rangle_t +  \nonumber \\
&&  3\langle m_1^{p-1}m_2^{p-1}q_{34}^{p-1} \rangle_t\Big],
\label{m1m2}
\end{eqnarray}
\begin{eqnarray}
&& \frac{d\langle m_1^{2(p-1)}m_2^{p-1} \rangle_t}{dt} = \\
\nonumber && 2\tilde{\beta}\Big[ \langle m_1^{2(p-1)}m_2^{2(p-1)}
\rangle_t\Big] +\tilde{\beta}\Big[\mbox{unfilled terms}\Big] +
O(\theta^2).
\end{eqnarray}
Integrating back in t and neglecting higher order terms we have
\begin{equation}
\langle m_1^{2(p-1)}m_2^{p-1} \rangle_t =
\tilde{\beta}\Big[\langle m_1^{2(p-1)}m_2^{2(p-1)} \rangle\Big]t,
\end{equation}
and we can write
\begin{eqnarray}
&& \langle m_1^{p-1}m_2^{p-1} \rangle_t = \\ \nonumber &&
\tilde{\beta}\theta\langle m_1^{p-1}m_2^{p-1}q_{12}^{p-1} \rangle
t + \tilde{\beta}^2\langle m_1^{2(p-1)}m_2^{2(p-1)} \rangle t^2.
\end{eqnarray}
Let us take a look now at the other overlap $\langle
m_1^{p-1}q_{12}^{p-1} \rangle_t$:
\begin{eqnarray}\label{m1m2}
\frac{d\langle m_1^{p-1}q_{12}^{p-1} \rangle_t}{dt} &=&
\tilde{\beta}\Big[
\langle m_1^{2(p-1)}q_{12}^{p-1} \rangle_t - \langle m_1^{p-1}m_2^{p-1}q_{12}^{p-1} \rangle_t  \nonumber \\
&-& 2\langle m_1^{p-1}m_2^{p-1}m_3^{p-1}q_{12}^{p-1}
\rangle_t\Big] + O(\theta^2),
\end{eqnarray}
that gives
\begin{equation}
\langle m_1^{p-1}q_{12}^{p-1} \rangle_t = \tilde{\beta}\langle
m_1^{p-1}m_2^{p-1}q_{12}^{p-1} \rangle t + O(\theta^2).
\end{equation}
\medskip
\newline
At this point we can write for $\langle
m_1^{p-1}\rangle_{t,\tilde{\alpha}}$ (and consequently for
$\langle q_{12}^{p-1}\rangle_{t,\tilde{\alpha}}$)
\begin{eqnarray}\nonumber
\langle m_1^{p-1}\rangle_{t,\tilde{\alpha}} &=&
\tilde{\beta}\langle m_1^{2(p-1)}\rangle t -
\frac{\tilde{\beta}^3}{3}\langle m_1^{2(p-1)}m_2^{2(p-1)}\rangle
t^3 \\ \nonumber &-&
\tilde{\beta}^2\theta\langle m_1^{p-1}m_2^{p-1}q_{12}^{p-1}\rangle t^2 + O(\theta^3), \nonumber \\
\langle q_{12}^{p-1}\rangle_{t,\tilde{\alpha}} &=&
\tilde{\beta}\theta\langle q_{12}^{2(p-1)}\rangle t +
\tilde{\beta}^2\langle m_1^{p-1}m_2^{p-1}q_{12}^{p-1}\rangle t^2 +
O(\theta^3).\nonumber
\end{eqnarray}
With these relations, eq. (\ref{Psit2}) becomes
\begin{eqnarray}\nonumber
&& \frac{d}{dt}\Psi_N(\alpha,\beta,t) = 2\alpha^{p-1}\ln\cosh\beta
+ \tilde{\beta}^2\langle m_1^{2(p-1)}\rangle t  \\ \nonumber && \
\ \ \  -\frac{\tilde{\beta}^2\theta^2}{2}\langle
q_{12}^{2(p-1)}\rangle t - \frac{3\tilde{\beta}^3\theta}{2}
\langle m_1^{p-1}m_2^{p-1}q_{12}^{p-1}\rangle t^2 \\ \nonumber &&
\ \ \ \  - \frac{\tilde{\beta}^4}{3}\langle
m_1^{2(p-1)}m_2^{2(p-1)}\rangle t^3 + O(\theta^5),\nonumber
\end{eqnarray}
which  ultimately allows us to write an iterated expressions for
$\Psi$ evaluated at $t=1$
\begin{eqnarray}
&& \Psi_N(\alpha,\beta,1) = \\ \nonumber &&
2\alpha^{p-1}\ln\cosh\beta + \frac{\tilde{\beta}^2}{2}\langle
m_1^{2(p-1)}\rangle  -
\frac{\tilde{\beta}^2\theta^2}{4}\langle q_{12}^{2(p-1)}\rangle - \\
&& \frac{\tilde{\beta}^3\theta}{2} \langle
m_1^{p-1}m_2^{p-1}q_{12}^{p-1}\rangle -
\frac{\tilde{\beta}^4}{12}\langle m_1^{2(p-1)}m_2^{2(p-1)}\rangle
t^3 + O(\theta^5). \nonumber
\end{eqnarray}
Overall the result we were looking for, a polynomial form of the
free energy, reads off as
\begin{eqnarray}\label{maino}
A(\alpha,\beta) &=& \ln2 \:+\: \alpha^{p-1}\ln\cosh\beta +\\
&+& \frac{\beta'}{2}\Big(\beta'\langle m^{2(p-1)}\rangle - \langle
m^{p}\rangle\Big) + \\ \nonumber &+&
\frac{\beta'\theta}{4}\Big(\beta'\theta\langle
q_{12}^{2(p-1)}\rangle - \langle q_{12}^{p}\rangle\Big) +
O(\theta^5). \nonumber
\end{eqnarray}

Now, several conclusions can be addressed from the expression
(\ref{maino}):
\newline
\begin{remark}
At first let us note that, by constraining the interaction to be
pairwise, critical behavior should arise \cite{landau}.
Coherently, we see that for $p=2$ we can write the free energy
expansion as
$$A(\alpha,\beta)_{p=2} = \ln 2 + \alpha
\ln\cosh(\beta) - \frac{\beta'}{2}(1-\beta')\langle m^2 \rangle -
\frac{\beta' \theta}{4} \langle q_2^2 \rangle,$$ which coincides
with the one of the diluted ferromagnet \cite{ABC} and
displays criticality at $2\alpha\theta=1$, where the coefficient
of the second order term vanishes, in agreement with previous
results\cite{ABC}.
\end{remark}
\begin{remark}
The free energy density of the fully connected $p$-spin model
is \cite{abarra3} $A(\beta')= \ln 2 + \ln\cosh(\beta m^{p-1}) -
(\beta/2)m^p$, which coincides with the expansion (\ref{maino}) in
the limit of $\alpha \to \infty$ and $\beta \to 0$ with $\beta' =
2(p-1) \alpha^{p-1}\theta$.
\end{remark}
\begin{remark}
It is worth noting that the connectivity no longer plays a linear
role in contributing to the free energy density, as it does happen
for the diluted two body models \cite{ABC,gt2}, but, in complete
generality as $p-1$. This is interesting in social networks,
where, for high values of coordination number it may be
interesting developing strategies with more than one exchange
\cite{barkema}.
\end{remark}
\begin{remark}
As from the numerics discussed in Section $2$, the effective $p$
turns out to be $2.15$; to check for consistency with the analytic
results, we note that close to criticality ($p=2$), the
temperature for the phase transition is given by $\beta_c =
\tanh^{-1}(1/2\alpha^{p-1})= \tanh^{-1}(1/2\alpha)$. Now, for $p=2.15$ this expression becomes $\beta_c \sim
\tanh^{-1}(1/2\alpha^{p-1})= \tanh^{-1}(1/2\alpha^{1.15})$: The
ratio among the two expressions, when evaluated for $\alpha=10$ gets approximately $1.4$, in agreement with data depicted in Fig.~$1$.
\end{remark}

\subsection{Numerics}\label{sec:numerics}

We now analyze the system described in this section, from the
numerical point of view by performing extensive Monte Carlo
simulations. Within this approach it is more convenient to use the
second Hamiltonian introduced (see eq.(\ref{parag})):
\begin{equation}\label{my_hamiltonian}
\hat{H}_N(\sigma)=-\sum_{i_i}^{N}
\sigma_{i_1}\sum_{i_2<i_3<...<i_p=1}^{N} A_{i_1,...,i_p}
\sigma_{i_{2}}\sigma_{i_{3}}...\sigma_{i_{p}}.
\end{equation}
The product between the elements of the adjacency tensor ensures
that the $p-1$ spins considered in the second sum are joined by a
link with $i_1$.
\newline
The evolution of the magnetic system is realized by means of a
single spin flip dynamics based on the Metropolis algorithm
\cite{barkema}. At each time step a spin is randomly extracted and
updated whenever its coordination number is larger than $p-1$. For
$\alpha$ large enough (at least above the percolation threshold,
as  obviously it is the case for the results found previously) and
$p=3,4$ this condition is generally verified. The updating
procedure for a spin $\sigma_i$ works as follows: Firstly we
calculate the energy variation $\Delta E_i$ due to a possible spin
flip, which for $p=3$ and $p=4$ reads respectively
\begin{eqnarray}
\Delta E_i &=& 2 \sigma_i \sum_{j<k=1}^{N} A_{i,j}A_{i,k}
\sigma_{j}\sigma_{k},
\\
\Delta E_i &=& 2 \sigma_i \sum_{j<k<w=1}^{N} A_{i,j}A_{i,k}A_{i,w}
\sigma_{j}\sigma_{k}\sigma_{w}.
\end{eqnarray}
Now, if $\Delta E_i <0$, the spin-flip $\sigma_i \rightarrow -
\sigma_i$ is realized with probability $1$, otherwise it is
realized with probability $e^{-\beta \Delta E}$.

The case $p=3$ has been studied in details and some insight is
provided also for the case $p=4$, while  for $p=2$ we refer to
\cite{ABC}. Our investigations concern two main issues:\\
- the existence of a phase transition and its nature\\
- the existence of a proper scaling for the temperature as the
parameter $\alpha$ is tuned.

As for the first point, we measured the so-called Binder cumulants
defined as follows:
\begin{equation}
G_N(T) \equiv 1 - \frac{\langle m^4 \rangle_N}{3\langle m^2
\rangle_N^2},
\end{equation}
where $\langle \cdot \rangle_N$ indicates the average obtained for
a system of size $N$ \cite{binder}. The study of Binder cumulants
is particularly useful to locate and  catalogue the phase
transition. In fact, in the case of continuous phase transitions,
$G_N(T)$ takes a universal positive value at the critical point
$T_c$, namely all the curves obtained for different system sizes
$N$ cross each other. On the other hand, for a first-order
transition $G_N(T)$ exhibits a minimum at $T_{min}$, whose
magnitude diverges as $N$. Moreover, a crossing point at
$T_{cross}$ can be as well detected when curves pertaining to
different sizes $N$ are considered \cite{vollmayr}. Now, $T_{min}$
and $T_{cross}$ scale as $T_{min}-T_c \propto N^{-1}$ and
$T_{cross}-T_c \propto N^{-2}$, respectively.

\begin{figure}[tb]
\bc
\includegraphics[height=60mm]{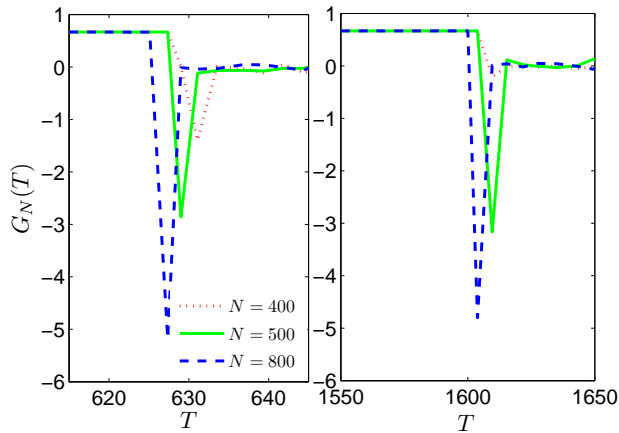}
\caption{\label{fig:Binder} Binder cumulants $G_L(T)$ for systems
of different size $N$, as shown in the legend, and connectivity
$\bar\alpha=50$ (left panel) and $\bar\alpha=80$ (right panel).}
\ec
\end{figure}

In Fig.~\ref{fig:Binder} we show data for $G_N(T)$ obtained for
systems of different sizes ($N=400$, $N=500$, and $N=800$) but
equal connectivity ($\alpha=50$ and $\alpha=80$, respectively) as
a function of the temperature $T$. The existence of a minimum is
clear and it occurs for $T \approx 625$ and $T \approx 1600$.
Similar results are found also for $p=4$ and they all highlight
the existence of a first-order phase transition at a temperature
which depends on the connectivity $\alpha$.

In order to deepen the role of connectivity in the evolution of
the system we measure the macroscopic observable $\langle m
\rangle$ and its (normalized) fluctuations $\langle m^2 \rangle -
\langle m \rangle^2$, studying their dependence on the temperature
$\beta$ and on the dilution $\alpha$. Data for different choices
of size and dilution are shown in Figure \ref{fig:p3}.

The profile of the magnetization, with an abrupt jump, and the
correspondent peak found for its fluctuations confirm the
existence of a first order phase transition at a well defined
temperature $T_c$ whose value depends on the dilution $\alpha$.
More precisely, by properly normalizing the temperature in
agreement with analytical results, namely $\tilde\beta \equiv
\beta \; \bar\alpha^{p-1}$ we found a very good collapse of all
the curves considered. Hence, we can confirm that the temperature
scales like $\alpha^{p-1}$. Moreover our data provide a very clear
hint suggesting that the critical temperature can be written as $T
= \alpha^{p-1}/4$.

\begin{figure}[tb]\bc
\includegraphics[height=60mm]{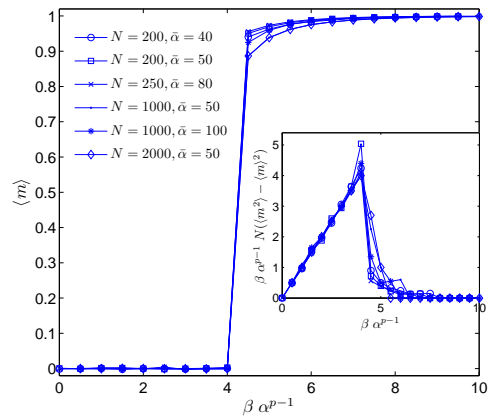}
\caption{\label{fig:p3}Magnetization (main figure) and its
normalized fluctuations (inset) for systems of different sizes and
different dilution as a function of $\beta \; \alpha^{p-1}$. The
collapse of all the curves provides a strong evidence for the
scaling of the temperature.} \ec
\end{figure}

\section{Conclusions}
We have seen in this work a numerical and analytical study of a
a diffusive strategy modeling group competition. The study is performed
on a standard random graph in a ferromagnetic mean field model.
The steady state equilibria show a {\it shifted-temperature} phenomena:
the Hamiltonian equilibria differ from the stationary one. We propose
an {\it effective} Hamiltonian description of the last by means of an
analytically continued random p-spin model which turns out to provide a
good approximation for the steady state for $p=2.15$. The analysis of
the free energy functional suggests moreover that the connectivity
gives a non linear contribution to the equilibrium state. This open the possibility
to consider extended models in which the connectivity is a dynamical
variable for the system and is chosen thermodynamically to maximize
stability. We plan to return on that topic in future works.

\section{Appendix: Analytical proofs}

In this section the proofs of al the Theorems and the Proposition
$1$ are reported.

\bigskip

\textbf{Proof of Theorem \ref{primolevi}}
\newline
Bridging a system made of by $N+1$ spins with one made of by $N$
spins implies the definition of rescaled $\gamma, \alpha$
parameters, accordingly to \cite{ABC,BCC}
\begin{eqnarray}
\tilde{\gamma} &=& \gamma \frac{N}{N+1} \qquad
\stackrel{N\rightarrow\infty}{\longrightarrow} \quad \gamma \\
\tilde{\alpha} &=& \alpha \Big[\frac{N}{N+1}\Big]^{\frac{1}{p-1}}
\qquad \stackrel{N\rightarrow\infty}{\longrightarrow} \quad
\alpha.
\end{eqnarray}
We have, in distribution, the Hamiltonian of a system made of
$N+1$ particles writable as
\begin{eqnarray}\label{Hsplit}
H_{N+1}(\sigma,\gamma) &=& -\sum_{\nu=1}^{k_{\gamma (N+1)}}
\sigma_{i_{\nu}^1}\sigma_{i_{\nu}^2}...\sigma_{i_{\nu}^p}\,\sim\,
\\ \nonumber &-& \sum_{\nu=1}^{k_{\tilde{\gamma} N}}
\sigma_{i_{\nu}^1}\sigma_{i_{\nu}^2}...\sigma_{i_{\nu}^p} -
\sum_{\nu=1}^{k_{2\tilde{\gamma}}}
\sigma_{i_{\nu}^1}\sigma_{i_{\nu}^2}...\sigma_{i_{\nu}^{p-1}}\sigma_{N+1},
\end{eqnarray}
that we may rewrite as \be H_{N+1}(\sigma,\gamma) =
H_N(\sigma,\tilde{\gamma}) + \hat{H}_N(\sigma,2\tilde{\gamma}).
\ee
\medskip
\newline
Following the above decomposition, let us consider the partition
function of the same $N+1$ spin model and let us introduce the
gauge transformation $\sigma_{i} \rightarrow \sigma_i\sigma_{N+1}$
which is a symmetry of the Hamiltonian known as
\textit{spin-flip}.
\begin{eqnarray}
Z_{N+1}(\gamma,\beta) &\sim& \sum_{\{\sigma_{N+1}\}} e^{-\beta
H_{N}(\sigma,\tilde{\gamma}) -
\beta\hat{H}_{N}(\sigma,\tilde{\gamma})\sigma_{N+1}} = \label{T11}
\\ \nonumber &=& \sum_{\{\sigma_{N+1}\}}
e^{\beta H_{N}(\sigma,\tilde{\gamma}) +
\beta\sum_{\nu=1}^{k_{2\tilde{\gamma }}}
\sigma_{i_{\nu}^1}...\sigma_{i_{\nu}^{p-1}}\sigma_{N+1}} =
\label{T12} \\ \nonumber &=& 2 \sum_{\{\sigma_{N}\}}
e^{\beta\sum_{\nu=1}^{k_{\tilde{\gamma}N}}
\sigma_{i_{\nu}^1}...\sigma_{i_{\nu}^p} +
\beta\sum_{\nu=1}^{k_{2\tilde{\gamma }}} \sigma_{i_{\nu}^1}...\sigma_{i_{\nu}^{p-1}}} = \label{T13} \\
&=& 2 Z_N(\tilde{\gamma},\beta)\tilde{\omega}(e^{-\beta
\hat{H}_{N}}),\nonumber
\end{eqnarray}
where  the new Boltzmann state $\tilde{\omega}$, and its
replicated $\tilde{\Omega}$, are introduced as \begin{eqnarray}
\tilde{\omega}(g(\sigma)) &=&
\frac{\sum_{\{\sigma_{N}\}}g(\sigma)e^{-\beta
H_N(\tilde{\gamma},\sigma)}} {\sum_{\{\sigma_{N}\}}e^{-\beta
H_N(\tilde{\gamma},\sigma)}}, \\  \qquad \tilde{\Omega}(g(\sigma))
&=& \prod_i\tilde{\omega}^{(i)}(g(\sigma^{(i)})). \end{eqnarray}
\newline
To continue the proof we now take the logarithm of both sides of
the last expression in eq. (\ref{T11}), apply the expectation
$\textbf{E}$ and subtract the quantity $\textbf{E}[\ln
Z_{N+1}(\tilde{\gamma},\beta)]$. We obtain
\begin{eqnarray}\nonumber
&& \textbf{E}[\ln Z_{N+1}(\gamma,\beta)] - \textbf{E}[\ln
Z_{N+1}(\tilde{\gamma},\beta)]=  \\  && \ln2 - \textbf{E}[\ln
\frac{Z_{N+1}(\gamma,\beta)}{Z_N(\tilde{\gamma},\beta)}] +
\Psi_N(\tilde{\gamma},\beta,1),
\end{eqnarray}
The left hand side gives
\begin{eqnarray}
\textbf{E}[\ln Z_{N+1}(\gamma,\beta)] &-& \textbf{E}[\ln
Z_{N+1}(\tilde{\gamma},\beta)] = \\ \nonumber &=&
(\gamma - \tilde{\gamma})\frac{d}{d\gamma}\textbf{E}[\ln Z_{N+1}(\gamma,\beta)]|_{\gamma=\tilde{\gamma}} = \nonumber \\
&=& \gamma\frac{1}{N+1}\frac{d}{d\gamma}\textbf{E}[\ln Z_{N+1}(\gamma,\beta)]|_{\gamma=\tilde{\gamma}} = \nonumber \\
&=& \gamma \frac{d}{d\gamma}A_{N+1}(\gamma,\beta).
\end{eqnarray}

Considering the $\alpha$ dependence of $\gamma$, we have
$$
\partial_{\gamma} \propto \frac{1}{(p-1)\alpha^{p-2}}\partial_{\alpha} \quad
\Rightarrow \quad \gamma\frac{d}{d\gamma}A \propto
\frac{\alpha}{p-1}\frac{d}{d\alpha}A,
$$
where the symbol $\propto$ instead of $=$ reflects the
arbitrariness by which we include  the $p!$ term, multiplying
$\alpha$, inside the definition of $\gamma$, or directly in
$\alpha$.
\newline
Performing now the thermodynamic limit, we see that at the right
hand side we have \be \lim_{N \to \infty}\textbf{E}[\ln
\frac{Z_{N+1}(\alpha,\beta)}{Z_N(\tilde{\alpha},\beta)}]
\longrightarrow A(\alpha,\beta) \ee and the theorem is proved
$\Box$.

\bigskip

\textbf{Proofs of Theorems
\ref{ciccia},\ref{saturabili},\ref{saturi}}
\newline
In this sketch we are going to show how to get Theorem
(\ref{ciccia}) in some details; It automatically has as a
corollary Theorem (\ref{saturabili}) which ultimately gives, as a
simple consequence when applied on filled monomials,
Theorem(\ref{saturi}).
\newline
Let us assume for a generic overlap correlation function $Q$, of
$s$ replicas, the following representation
$$
Q = \prod_{a=1}^s\sum_{i_l^a}\prod_{l=1}^{n^a}\sigma_{i_l^a}^a
I(\{i_l^a \})
$$
where $a$ labels the replicas, the internal product takes into
account the spins (labeled by $l$) which contribute to the a-part
of the overlap $q_{a,a'}$ and runs to the number of time that the
replica $a$ appears in $Q$. The external product takes into
account all the contributions of the internal one and the $I$
factor fixes the constraints among different replicas in $Q$; so,
for example, $Q=q_{12}q_{23}$ can be decomposed in this form
noting that $s=3$, $n^1=1,n^2=2$,
$I=N^{-2}\delta_{i_1^1,i_1^3}\delta_{i_1^2,i_2^3}$, where the
$\delta$ functions fixes the links between replicas $1,3
\rightarrow q_{1,3}$ and $2,3 \rightarrow q_{2,3}$. The averaged
overlap correlation function is
$$
\langle Q \rangle_t = \mathbf{E}\sum_{i_l^a}I(\{i_l^a
\})\prod_{a=1}^s \omega_{t}(\prod_{l=1}^{n^a}\sigma_{i_l^a}^a).
$$
Now if $Q$ is a fillable polynomial, and we evaluate it at $t=1$,
let us decompose it, using the factorization of the $\omega$ state
on different replica, as
$$ \langle Q \rangle_t = \mathbf{E}\sum_{i_l^a,i_l^b}I(\{i_l^a \}, \{i_l^b \})\prod_{a=1}^u
\omega_a ( \prod_{l=1}^{n^a}\sigma_{i_l^a}^a) \prod_{b=u}^s
\omega_b ( \prod_{l=1}^{n^b}\sigma_{i_l^b}^b),
$$
where $u$ stands for the number of the unfilled replicas inside
the expression of $Q$. So we split the measure $\Omega$ into two
different subset $\omega_{a}$ and $\omega_{b}$: in this way the
replica belonging to the $b$ subset are always in even number,
while the ones in the $a$ subset are always odds. Applying the
gauge $\sigma_i^a \rightarrow \sigma_i^a\sigma_{N+1}^a, \forall i
\in (1,N)$ the even measure is unaffected by this transformation
$(\sigma_{N+1}^{2n} \equiv 1)$ while the odd measure takes a
$\sigma_{N+1}$ inside the Boltzmann measure. \begin{eqnarray} &&
\langle Q \rangle = \\ \nonumber && \sum_{i_l^a,i_l^b}I(\{i_l^a
\}, \{i_l^b \}) \prod_{a=1}^u \omega ( \sigma_{N+1}^a
\prod_{l=1}^{n^a}\sigma_{i_l^a}^a) \prod_{b=u}^s \omega (
\sigma_{N+1}^b\prod_{l=1}^{n^b}\sigma_{i_l^b}^b). \end{eqnarray}
At the end we can replace in the last expression the index $N+1$
of $\sigma_{N+1}$ by $k$ for any $k \neq \{ i_l^a \}$ and multiply
by one as $1=N^{-1}\sum_{k=0}^N$. Up to orders $O(1/N)$, which go
to zero in the thermodynamic limit, we have the proof.
\medskip
\newline
It is now immediate to understand that Theorem (\ref{ciccia}) on a
fillable overlap monomial has the effect of multiplying it by its
missing part to be filled (Theorem \ref{saturabili}), while it has
no effect if the overlap monomial is already filled (Theorem
\ref{saturi}). $\Box$

\bigskip

\textbf{Proof of Proposition \ref{stream}}
\newline
The proof works by direct calculation:
\begin{eqnarray}
&& \frac{\partial\langle F_s \rangle_{t,\tilde{\alpha}}}{\partial
t} = \\ \nonumber && \frac{\partial \textbf{E}}{\partial t}  \Big[
\frac{\sum_{\{\sigma\}}F_s
e^{\sum_{a=1}^s(\beta\sum_{\nu=1}^{k_{\tilde{\gamma} N}}
\sigma_{i_{\nu}^1}^a...\sigma_{i_{\nu}^p}^a +
\beta\sum_{\nu=1}^{k_{2\tilde{\gamma}t}}
\sigma_{i_{\nu}^1}^a...\sigma_{i_{\nu}^{p-1}}^a)}}
{\sum_{\{\sigma\}}
e^{\sum_{a=1}^s(\beta\sum_{\nu=1}^{k_{\tilde{\gamma} N}}
\sigma_{i_{\nu}^1}^a...\sigma_{i_{\nu}^p}^a + \beta
\sum_{\nu=1}^{k_{2\tilde{\gamma}t}}
\sigma_{i_{\nu}^1}^a...\sigma_{i_{\nu}^{p-1}}^a)}}\Big] =
\\ \nonumber && 2\tilde{\alpha}^{p-1} \textbf{E}\Big[
\frac{\tilde{\Omega}_t (F_s
e^{\sum_{a=1}^s(\beta\sigma_{i_{0}^1}^a...\sigma_{i_{0}^{p-1}}^a)})}
{\tilde{\Omega}_t(e^{\sum_{a=1}^s(\beta\sigma_{i_{0}^1}^a...\sigma_{i_{0}^{p-1}}^a)})}\Big]
- 2\tilde{\alpha}^{p-1}\langle F_s \rangle_{t,\tilde{\alpha}} = \\
\nonumber &&  2\tilde{\alpha} \textbf{E}\Big[
\frac{\tilde{\Omega}_t (F_s \Pi_{a=1}^{s}(\cosh\beta +
\sigma_{i_{0}^1}^a...\sigma_{i_{0}^{p-1}}^a\sinh\beta))}
{\tilde{\Omega}_t (\Pi_{a=1}^{s}(\cosh\beta +
\sigma_{i_{0}^1}^a...\sigma_{i_{0}^{p-1}}^a\sinh\beta))}\Big]- \\
\nonumber &&
2\tilde{\alpha}^{p-1}\langle F_s \rangle_{t,\tilde{\alpha}} = \\
\nonumber &&   2\tilde{\alpha}^{p-1} (\textbf{E}\Big[
\frac{\tilde{\Omega}_t (F_s \Pi_{a=1}^{s}(1 +
\sigma_{i_{0}^1}^a...\sigma_{i_{0}^{p-1}}^a\theta))} {(1 +
\tilde{\omega}_t(\sigma_{i_{0}^1}^a...\sigma_{i_{0}^{p-1}}^a)\theta)^s}\Big]
- \langle F_s \rangle_{t,\tilde{\alpha}}),
\end{eqnarray}
Now noting that
\begin{eqnarray} \nonumber
\Pi_{a=1}^{s}(1 &+&
\sigma_{i_{0}^1}^a...\sigma_{i_{0}^{p-1}}^a\theta) = 1 +
\sum_{a=1}^{s}\sigma_{i_{0}^1}^a...\sigma_{i_{0}^{p-1}}^a\theta
\\ \nonumber &+& \sum_{a<b}^{1,s}\sigma_{i_{0}^1}^a...\sigma_{i_{0}^{p-1}}^a
\sigma_{i_{0}^1}^b...\sigma_{i_{0}^{p-1}}^b\theta^2
+ ...\nonumber \\
\frac{1}{(1 + \tilde{\omega}_t \theta)^s} &=& 1 -
s\tilde{\omega}_t \theta + \frac{s(s+1)}{2!}\tilde{\omega}_t^2
\theta^2 + ... \nonumber
\end{eqnarray}
\medskip
we obtain
\begin{eqnarray}
\frac{\partial\langle F_s \rangle_{t,\tilde{\alpha}}}{\partial t}
&=& 2\tilde{\alpha}^{p-1} \Big(\textbf{E}\Big[ \tilde{\Omega}_t
\Big(F_s(1 +
\sum_{a=1}^{s}\sigma_{i_{0}^1}^a...\sigma_{i_{0}^{p-1}}^a\theta +
\\ \nonumber &+& \sum_{a<b}^{1,s}\sigma_{i_{0}^1}^a...\sigma_{i_{0}^{p-1}}^a
\sigma_{i_{0}^1}^b...\sigma_{i_{0}^{p-1}}^b\theta^2
+ ...)\Big) \times \nonumber \\
&\times& \Big(1 - s\tilde{\omega}_t \theta +
\frac{s(s+1)}{2!}\tilde{\omega}_t^2 \theta^2 + ...\Big)\Big] -
\langle F_s \rangle_{t,\tilde{\alpha}}\Big), \nonumber
\end{eqnarray}
from which our thesis follows. $\Box$

\end{document}